\documentclass[aps,twocolumn,superscriptaddress]{revtex4}
\usepackage{graphics}
\usepackage[pdftex]{graphicx}
\usepackage{times}
\usepackage{color}
\begin{document}

\title{Spreading of cooperative behaviour across interdependent groups}

\author{Luo-Luo Jiang}
\email{jiangluoluo@gmail.com}
\affiliation{College of Physics and Electronic Information Engineering, Wenzhou University, 325035 Wenzhou, China}

\author{Matja{\v z} Perc}
\email{matjaz.perc@uni-mb.si}
\affiliation{Faculty of Natural Sciences and Mathematics, University of Maribor, Koro{\v s}ka cesta 160, SI-2000 Maribor, Slovenia}

\begin{abstract}
Recent empirical research has shown that links between groups reinforce individuals within groups to adopt cooperative behaviour. Moreover, links between networks may induce cascading failures, competitive percolation, or contribute to efficient transportation. Here we show that there in fact exists an intermediate fraction of links between groups that is optimal for the evolution of cooperation in the prisoner's dilemma game. We consider individual groups with regular, random, and scale-free topology, and study their different combinations to reveal that an intermediate interdependence optimally facilitates the spreading of cooperative behaviour between groups. Excessive between-group links simply unify the two groups and make them act as one, while too rare between-group links preclude a useful information flow between the two groups. Interestingly, we find that between-group links are more likely to connect two cooperators than in-group links, thus supporting the conclusion that they are of paramount importance.
\end{abstract}

\maketitle

According to a recent study by Apicella et al. \cite{apicella_n12}, social networks of the Hadza, a population of hunter-gatherers in Tanzania, have much in common with modernized social networks \cite{wasserman_94, christakis_09}. Moreover, and of direct relevance for the present study, Hadza camps exhibit high between-group and low within-group variation in public goods game donations, and the links are more likely between people who do cooperate than between those who do not. Authors of \cite{apicella_n12} go on to conclude that early humans may have formed ties with both kin and non-kin, based in part on their tendency to cooperate, and that thus social networks may have actually contributed to the emergence and facilitated the evolution of cooperation \cite{rand_pnas11, wang_j_pnas12}. Inspired by these results, we here address the relevance of the interdependence between groups for the evolution of cooperation by means of numerical simulations. We consider two groups of certain size, whereby members of one group are allowed to break one of their in-group links to connect with a member in the other group, i.e., to form a between-group link. This set-up is akin to previous studies that have addressed the evolution of cooperation on interdependent networks \cite{wang_z_epl12, wang_z_srep13, gomez-gardenes_srep12, gomez-gardenes_pre12, wang_b_jsm12, szolnoki_njp13}, but also different in that we consider between-group links to actually replace in-group links, and also by considering between-group links as fully equivalent to in-group links in the sense that both payoff accumulation and strategy transfer across them are allowed.

The promotion of cooperation on networks in general is due to network reciprocity -- a phenomenon first reported by Nowak and May \cite{nowak_n92b}, who observed that on a square lattice cooperators can aggregate into compact clusters and so protect themselves against defectors when playing a prisoner's dilemma game. Network reciprocity quickly rose to prominence through a series of subsequent investigations on regular lattices and graphs \cite{lindgren_pd94, nowak_ijbc94, szabo_pre98, hauert_prslb01}, and even more so through studies of evolutionary games on small-world \cite{abramson_pre01, kim_bj_pre02, masuda_pla03, tomassini_pre06, fu_epjb07} and scale-free \cite{santos_prl05, santos_pnas06, gomez-gardenes_prl07, poncela_njp07, rong_pre07, masuda_prsb07, tomassini_ijmpc07, szolnoki_pa08, assenza_pre08, santos_n08, pena_pre09, poncela_pre11, brede_epl11, tanimoto_pre12, pinheiro_pone12} networks. Several recent reviews cover the topic in detail, both for pairwise social dilemmas, such as the prisoner's dilemma and the snowdrift game \cite{szabo_pr07, roca_plr09, perc_bs10}, as well as for evolutionary games that are governed by group interactions, such as the public goods game \cite{perc_jrsi13}. Based on evolutionary games, in particular on the ability of influential players to hinder the evolution of cooperation, also a network centrality measure has recently been introduced -- the so-called game centrality \cite{simko_pone13}. Despite its prominence, however, network reciprocity has also received a fair share of scepticism. Hauert and Doebeli \cite{hauert_n04} reported that spatial structure often inhibits the evolution of cooperation in the snowdrift game, while recent large-scale human experiments performed by Gracia-L{\'a}zaro et al. \cite{gracia-lazaro_srep12, gracia-lazaro_pnas12} revealed that network reciprocity may fail altogether.

\begin{figure}
\centering{\includegraphics[width = 5.5cm]{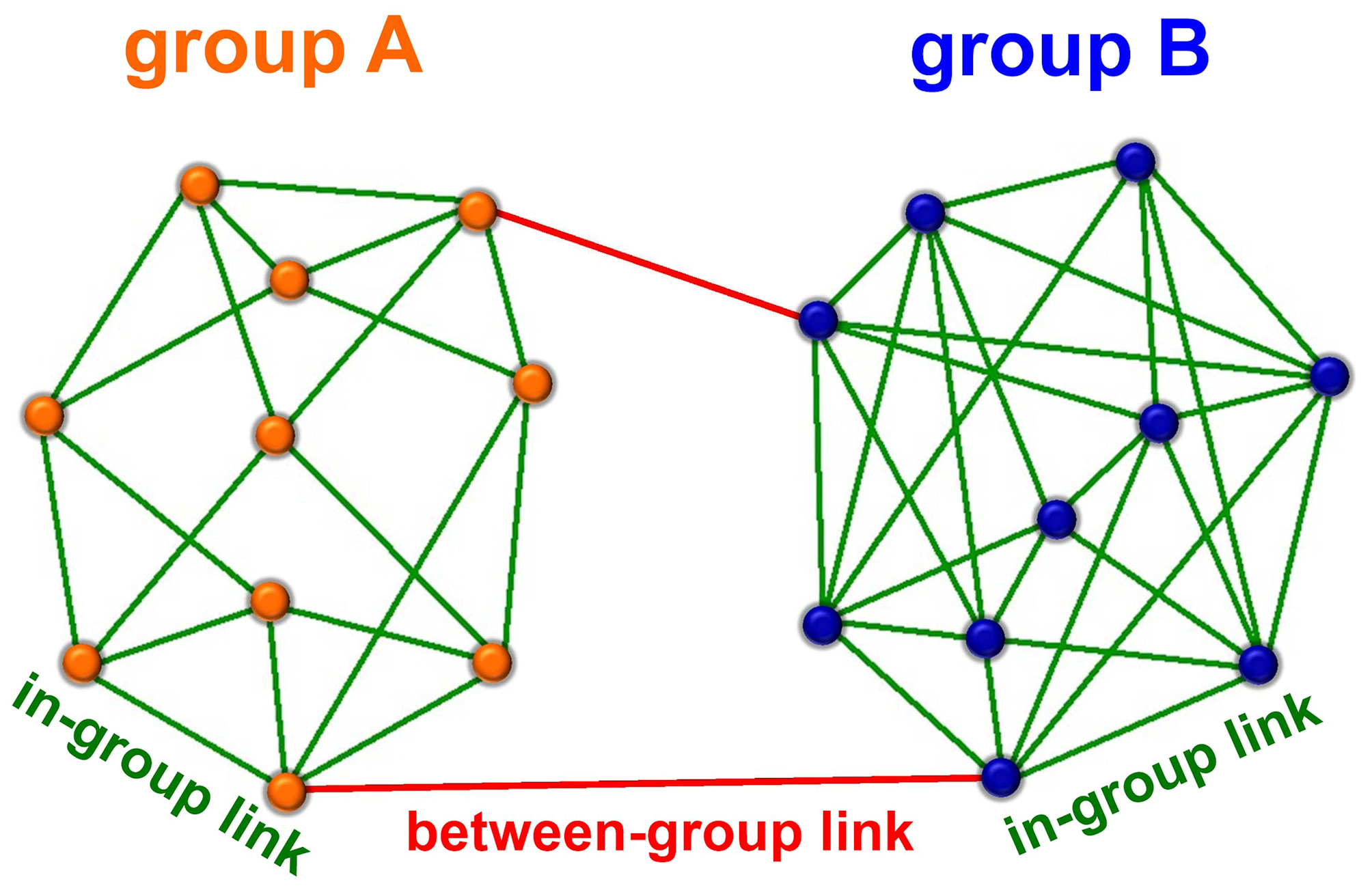}}
\caption{Schematic presentation of two groups that are connected by means of between-group links. Players in group A originally form a random network, each with degree four, while players in group B form a regular graph, each with degree six. Subsequently, each player in group B severs one of its in-group links (depicted green) with probability $p$ and uses it to form a between-group link (depicted red) with one randomly selected player from group A, the constrain being that no player in either groups is allowed to have more than one between-group link. In the example each group consists of $10$ players and $p=0.2$.}
\label{scheme}
\end{figure}

Apart from the evolution of cooperation on interdependent networks \cite{wang_z_epl12, wang_z_srep13, gomez-gardenes_srep12, gomez-gardenes_pre12, wang_b_jsm12, szolnoki_njp13}, showing that interdependence works predominantly in favour of the resolution of social dilemmas, previous research concerning interdependent networks has addressed cascading failures \cite{buldyrev_n10, li_w_prl12, parshani_pnas11, brummitt_pnas12}, competitive percolation \cite{parshani_prl10, nagler_np11}, transport \cite{morris_prl12}, diffusion \cite{gomez_prl13}, neuronal synchronization \cite{sun_xj_chaos11}, financial trading \cite{feng_pnas12}, as well as their robustness against attack and assortativity \cite{huang_xq_pre11, zhou_d_pre12}. Networks of networks have indeed captured the current attention of researchers across both social and natural sciences \cite{gao_jx_np12, havlin_pst12, helbing_n13, csermely_tde13}, and here we aim to extend their scope further to the evolution of cooperation among interdependent groups. The importance of groups in evolutionary games is well established, either through group competition \cite{bowles_s06}, group selection \cite{maynard_n64, wilson_ds_pnas75}, or the related multilevel selection \cite{traulsen_pnas06, szolnoki_njp09}.

In this paper, individual groups are represented by networks with different topology to account for different societal types, while links between groups are formed probabilistically by randomly choosing members in one group to break one of their existing in-group links to form a between-group link with a player from the other group. The set-up is depicted schematically in Fig.~\ref{scheme}. Once established, between-group links do not change during the course of evolution. All the players then engage in the prisoner's dilemma game that is characterized by the temptation to defect $T=b$, reward for mutual cooperation $R = 1$, and punishment $P$ as well as the sucker's payoff $S$ equalling $0$, whereby $1 < b \leq 2$ ensures a proper payoff ranking \cite{nowak_n92b}. The two main parameters to be considered in the Results section are the probability to establish between-group links $p$ and the temptation to defect $b$. For further details with regards to the studied evolutionary game and the set-up we refer to the Methods section. Largely independent of the topology of individual groups, we will show that intermediate interdependence optimally facilitates the spreading of cooperative behaviour between groups, and that between-group links are in fact crucial for the more favourable outcome and thus for the resolution of the prisoner's dilemma.

\section*{Results}

\begin{figure}
\centering{\includegraphics[width = 8.4cm]{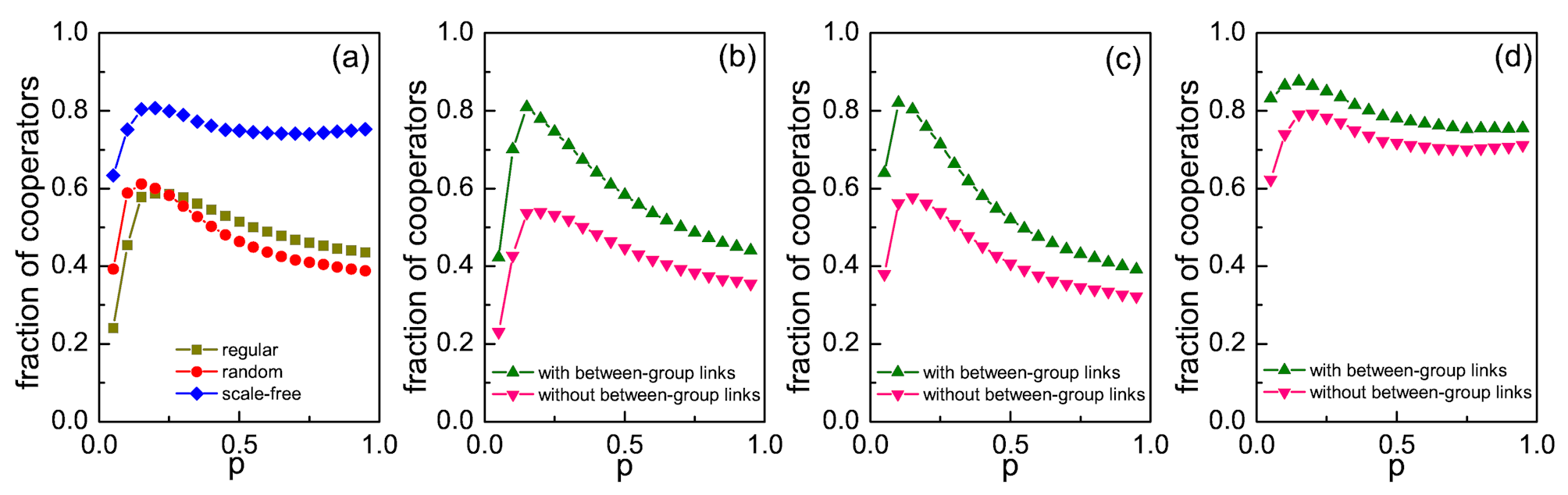}}
\caption{There is an intermediate fraction of between-group links at which the fraction of cooperators is maximal, regardless of the topologies of the two interdependent groups. Panel (a) depicts the overall fraction of cooperators in both groups in dependence on the probability to establish between-group links, as obtained for different topologies of group A (see legend). Panels (b), (c) and (d) depict the fraction of cooperators amongst players with and without between-group links separately (see legend), as obtained when group A has regular, random and scale-free topology, respectively. It can be observed that cooperation is significantly more likely amongst player that do have between-group links. Results were obtained using the temptation to defect $b=1.18$.}
\label{fraction}
\end{figure}

We begin by showing in Fig.~\ref{fraction}(a) the fraction of cooperators $f_C$ within both groups in dependence on the probability to form between-group links $p$ for different topologies of group A. Without loss of generality, the topology of group B is initially (before between-group links are formed) always regular. It can be observed that, regardless of the topology of group A, there exists an intermediate value of $p$ at which $f_C$ is maximal. The biggest rise compared to the base value obtained for independent groups ($p=0$) is obtained when the topology of group A is regular, followed by the random and scale-free topology. Especially for the late case the marginal improvement is expected, given that scale-free networks alone provide a very favourable environment for the evolution of cooperation \cite{santos_prl05, santos_pnas06}. Panels (b), (c) and (d) of Fig.~\ref{fraction} depict the fraction of cooperators separately for players with and without between-group links when group A has regular, random and scale-free topology, respectively (see figure legend). As for the overall fraction of cooperators, looking at $f_C$ separately for players with and without between-group links preserves the existence of an optimal values of $p$. In fact, $f_C$ peaks at roughly the same value of $p$ for players with and without between-group links, only that for the former the base-line value of $f_C$ is higher. It can thus be concluded that players with between-group links are more likely to cooperate than players without between-group links.

\begin{figure}
\centering{\includegraphics[width = 8.5cm]{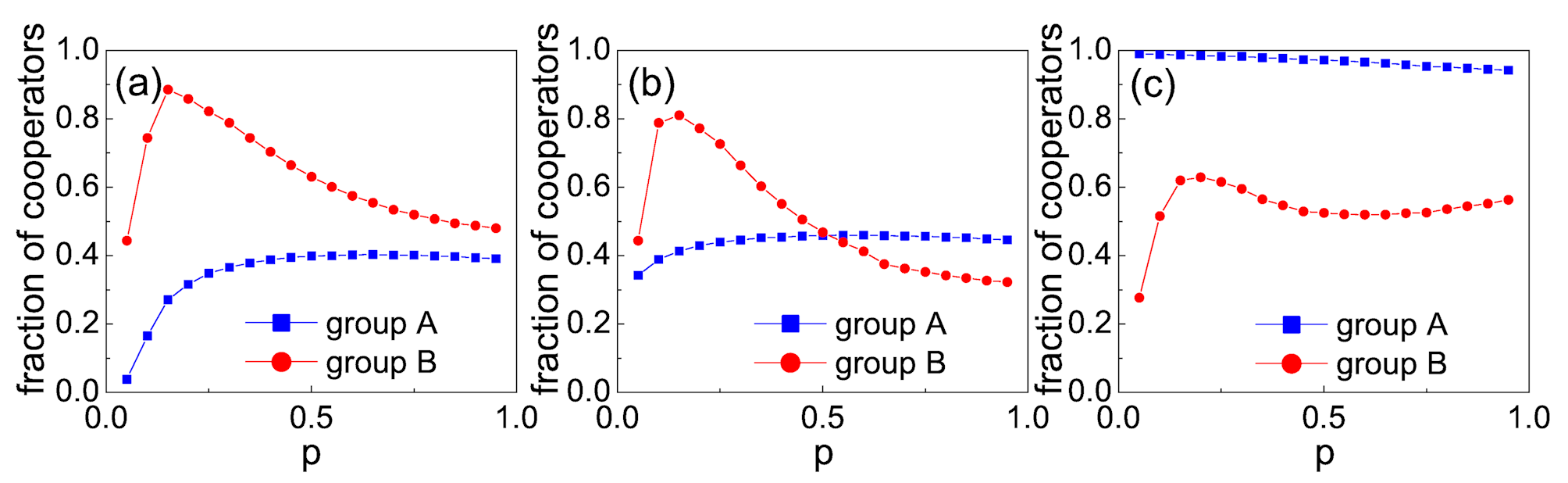}}
\caption{\label{groupfc} Cooperators are distributed unevenly between the two interdependent groups, depending on the topologies that govern the interactions in each individual group. Panels (a), (b) and (c) depict the fraction of cooperators in groups A and B separately (see legend), as obtained when group A has regular, random and scale-free topology, respectively. If group B is a regular graph, the dilution of links there has a more potent positive impact on cooperation (see \cite{wang_z_srep12, wang_z_pre12b} for related work) than the additional links introduced to group A. Players in group B in general always benefit from an optimal dilution, while players in group A benefit most from the additional links if initially they form a regular graph. If the topology of group A is random [as in panel (b)] or scale-free [as in panel (c)], the new between-group links stemming from players in group B have at most a marginal impact. As in Fig.~\ref{fraction}, results were obtained using the temptation to defect $b=1.18$.}
\end{figure}

Further supporting these observations are results presented in Fig.~\ref{groupfc}, which show the fraction of cooperators $f_C$ in dependence on $p$ separately for groups A and B (see figure legend). As in Fig.~\ref{fraction}, we consider group A with regular, random and scale-free topology in panels (a), (b) and (c), respectively, while the topology of group B is always regular. It can be observed that the bell-like shape of the $f_C$ versus $p$ dependence stems mainly from the group B, i.e., the group where players severe their in-group links to use them for forming between-group links. In group A, on the other hand, $f_C$ simply increases with increasing $p$ for the regular [panel (a)] and random [panel (b)] topology, or remains practically unchanged for the scale-free [panel (c)] topology. Since between-group links are added to the existing in-group links forming group A, the increase in $f_C$ can be understood as a consequence of the increase of heterogeneity of the interaction network of group A. Indeed, the positive role of heterogeneity (diversity) for the evolution of cooperation is firmly established, both for games governed by pairwise interactions \cite{santos_jtb12} as well as for games governed by group interactions \cite{perc_njp11}. In case group A is characterized by the scale-free topology, however, the addition of between-group links can hardly elevate the heterogeneity, and thus there $f_C$ remains practically constant regardless of the value of $p$ [see Fig.~\ref{groupfc}(c)]. The situation in group B is different, as there the existing in-group links are removed to be substituted by between-group links. Effectively the regular topology of group B becomes more and more diluted as $p$ increases, and indeed this may give rise to a bell-shaped outlay of $f_C$ versus $p$. The role of dilution, albeit by removing players rather than links, has recently been studied in \cite{wang_z_srep12, wang_z_pre12b}, and it was reported that such bell-shaped dependencies are indeed very much characteristic on diluted lattices. Notably, in Fig.~\ref{groupfc} the bells for group B differ since the links are not simply removed, but rather rewired to players that form group A. From there the support for cooperation differs depending on the topology of group A, which of course influences the outcome of the evolutionary process also in group B.

\begin{figure}
\centering{\includegraphics[width = 8.5cm]{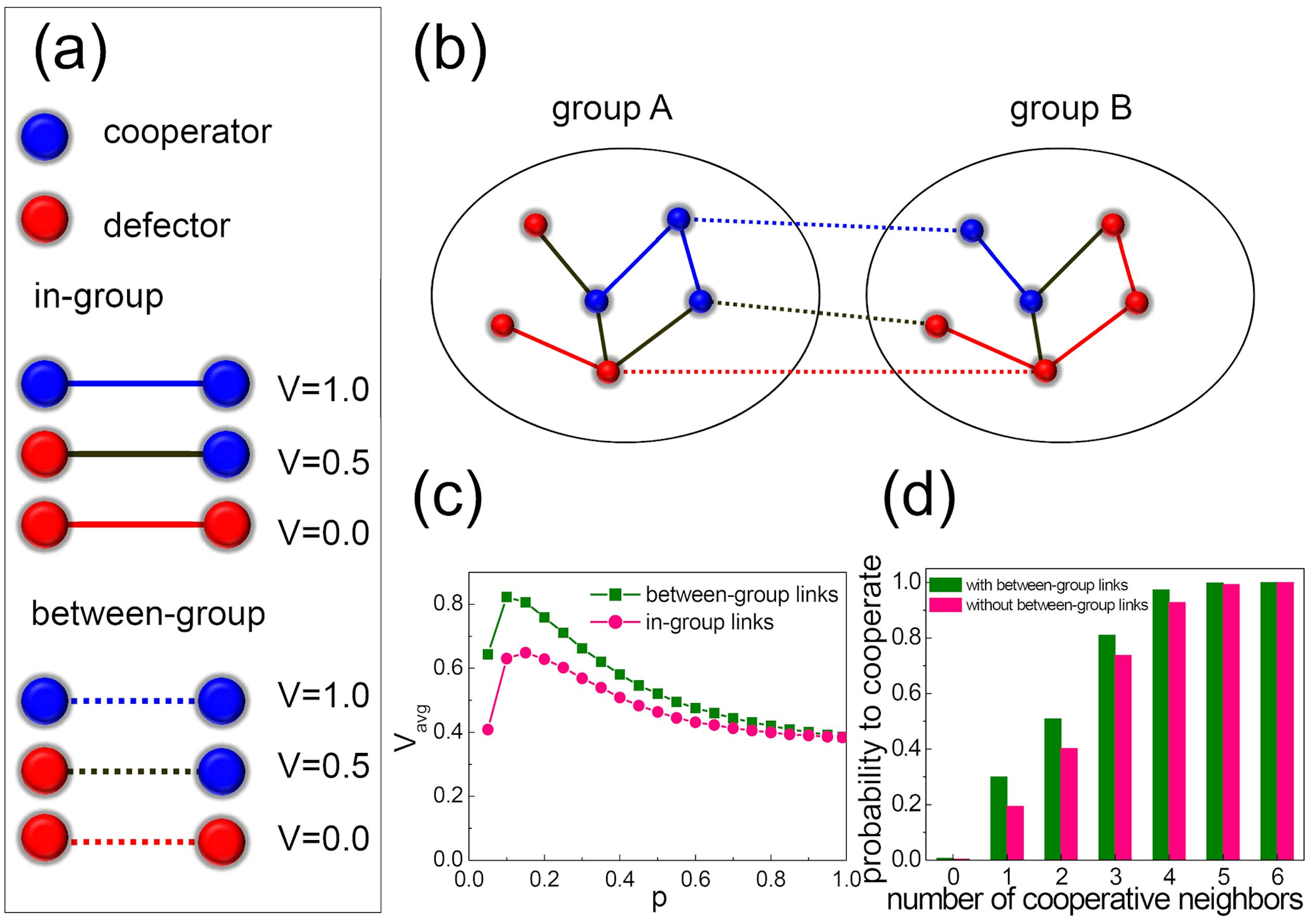}}
\caption{\label{explain} Players that share a between-group link are more likely to both cooperate than players that are connected by means of in-group links. Panels (a) and (b) explain and schematically depict the determination of $V$, which quantifies the cooperativity of each individual link. If a link connects two cooperators (defectors) $V=1$ ($V=0$), while cooperator-defector links yield $V=0.5$. The example in panel (b) thus yields $V_{avg}=5.5/12.0 \approx0.46$ for in-group links, and $V_{avg}=1.5/3.0=0.5$ for between-group links. Averaging $V$ separately over all in-group and between-group links under actual conditions yields results presented in panel (c), from where it follows that between-group links are more likely to connect two cooperators than in-group links, especially for low and intermediate values of $p$. This conclusion is further corroborated by results presented in panel (d), where we show the probability of a player to cooperate in the next round in dependence on the number of its cooperative neighbours. It can be observed that the probability is higher for players that have between-group links than for players without between-group links, especially if the player has only one, two, three of four cooperative neighbours. Results presented in panels (c) and (d) were obtained using random topology for group A and the temptation to defect $b=1.18$. For panel (d) we have used $p=0.15$.}
\end{figure}

From results presented in Figs.~\ref{fraction} and \ref{groupfc} it follows that links, especially between-group links, play a crucial role by the spreading of cooperative behaviour across interdependent groups. Intuitively, it can be argued that in-group links are responsible for the transmission of cooperation within each group, while between-group links help to spread cooperation  also past the boundaries of individual groups. But which links are more important, and which are more likely to contribute to an efficient spread of cooperative behaviour? To determine the role of the two types of links more accurately, we introduce $V$ as depicted schematically in Fig.~\ref{explain}(a). If a links connects two cooperators $V=1$. If a link connects two defectors $V=0$. And finally, if a link connects a cooperator and a defector $V=0.5$. In this way we obtain a proxy for the role links play, and by averaging the value of $V$ over all the in-group and between-group links in the system, as schematically depicted in Fig.~\ref{explain}(b), we obtain a better understanding of their relevance. Results presented in Fig.~\ref{explain}(c) show how $V_{avg}$ varies in dependence on $p$, separately for between-group and in-group links (see figure legend). Expectedly, given the definition of $V$, the dependence is bell-shaped with the maximum occurring at the same value of $p$ as the maximum of $f_C$ in Fig.~\ref{groupfc}(b) [where group A has random topology as used also in Fig.~\ref{explain}(c)]. More importantly, it can be observed that between-group links are more likely to connect cooperative pairs than in-group links. Thus, between-group links appear to be crucial for the spreading of cooperative behaviour across interdependent groups. Further adding weight to this statement are results presented in Fig.~\ref{explain}(d), where the probability to cooperate in the next round is depicted in dependence on the number of cooperative neighbours, separately for players with and without between-group links (see figure legend). It can be observed that the probability to cooperate is higher for players that have between-group links than it is for players without between-group links, especially if the player has only one, two, three of four cooperative neighbours. We thus conclude that between-group links are indeed more likely to link two cooperators than in-group links, and that they are in fact more crucial for the spreading of cooperative behaviour. These findings are in agreement with the recent empirical observations in the Hadza camps \cite{apicella_n12}, where it was observed that there is high between-group and low within-group variation in public goods game donations, and that the links between camps are more likely between people who do cooperate than between those who do not. The fact that players with more cooperative neighbours are in general more likely to cooperate in the next round, as can be observed in Fig.~\ref{explain}(d), also agrees with the observations of Traulsen et al. \cite{traulsen_pnas10}, who reported the same behaviour in the realm of human strategy updating in evolutionary games. What is remarkable in our case is that this probability is higher for players with between-group links than it is for players without between-group links.

\begin{figure}
\centering{\includegraphics[width = 8.5cm]{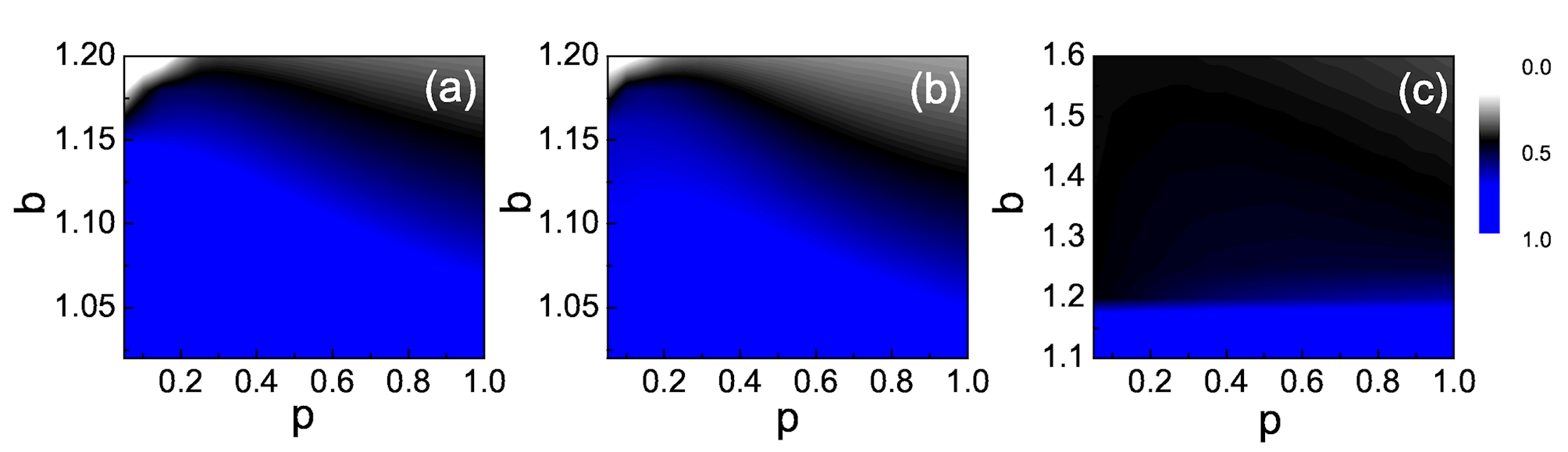}}
\caption{Optimal interdependence between two groups depends on the temptation to defect. Panels (a), (b) and (c) present the colour-encoded fraction of cooperators in both groups in dependence on the probability to establish between-group links $p$ and the temptation to defect $b$ when the topology of group A is regular, random and scale-free, respectively. If the topology of group A is regular, panel (a) reveals that there exists an optimal value of $p$ only for sufficiently large $b \geq 1.15$. If the topology of group A is random, there exists an optimal $p$ almost independently of $b$, although the optimal value of $p$ shifts towards lower values as $b$ increases. If group A has scale-free topology, however, the impact of $p$ is most illusive, since the scale-free topology alone strongly promotes the evolution of cooperation (see \cite{santos_prl05, santos_pnas06}).}
\label{colormap}
\end{figure}

Lastly, we present in Fig.~\ref{colormap} colour maps encoding the overall fraction of cooperators in both groups in dependence on $p$ and $b$, thus obtaining a more comprehensive insight as to the relevance of between-group links under differently severe social dilemma conditions. The outcome depends quite significantly on the topology of group A. If the topology is regular, as is the case in panel (a), there exists an optimal value of $p$ only for $b>1.15$, while for $b<1.15$ the impact of between-group links is predominantly negative. If the topology of group A is random, on the other hand, there exists an optimal $p$ irrespective of $b$, i.e., even if the conditions for cooperation are relatively favourable, although the optimal value of $p$ shifts towards lower values as $b$ increases. For group A having scale-free topology the impact is rather negligible regardless of $b$, as discussed already above when presenting results in Figs.~\ref{fraction} and \ref{groupfc}. In conclusion, these results show that significant advantages of group interdependence are to be expected only when the conditions for the evolution of cooperation are harsh, and when isolated groups alone are hardly able to keep defectors at bay. It is then that between-group links can do wonders in channelling cooperative behaviour from one group to another, and in doing so strengthening cooperative behaviour in each of them beyond the limits imposed by isolation.

\section*{Discussion}
Summarizing, we have studied the evolution of cooperation in the prisoner's dilemma game on interdependent groups having different interaction topology. We have shown that there exists an optimal probability for a player in one group to severe an in-group link and use this to establish a between-group link with one randomly selected player from another group. This conclusion is largely independent of the topology of the two groups, although it can be observed best if groups initially have either regular or random topology. If the initial topology is scale-free, the positive impact of the scale-free topology alone \cite{santos_prl05, santos_pnas06} precludes significant improvements in the level of cooperation that would be due to group interdependence. If the initial topology is regular or random, however, notable positive effects stem from the dilution of links in one group and the addition of new links to the other group, although the former effect in general proves to be stronger (for related work see \cite{wang_z_srep12, wang_z_pre12b}). We have also shown that players who have between-group links are more likely to cooperate than players without between-group links, and that between-group links are indeed more likely to link two cooperators than in-group links. These findings resonate with recent empirical observations in the Hadza camps \cite{apicella_n12}, where it was observed that these camps exhibit high between-group and low within-group variation in public goods game donations, and that the links between camps are more likely between people who do cooperate than between those who do not. Our findings also agree with previous theoretical research focusing on the evolution of cooperation on interdependent networks, where it has been shown, for example, that the coupling of the evolutionary dynamics in each of the two networks enhances the resilience of cooperation, and that this is intrinsically related to the non-trivial organization of cooperators across the interdependent layers \cite{gomez-gardenes_srep12}. It was also reported that biased utility functions suppress the feedback of individual success, which leads to a spontaneous separation of characteristic time scales on the two interdependent networks \cite{wang_z_epl12}. Consequently, cooperation is promoted because the aggressive invasion of defectors is more sensitive to the deceleration. Even if the utilities are not biased, cooperation can still be promoted by means of interdependent network reciprocity \cite{wang_z_srep13}, which however requires simultaneous formation of correlated cooperative clusters on both networks. Altogether, these results point to the fact that interdependence, be it between groups or other organizational entities, can be exploited effectively to resolve social dilemmas. Yet too much interdependence is not good either -- there must also be sufficient independence for the individual networks to remain functional if the evolution of cooperation in the other network goes wrong.

\section*{Methods}
Groups are initially constructed either with a regular topology where each player is connected to its $k$ nearest neighbours, or with a random topology where each player is also connected to $k$ other players, yet the latter are selected randomly from within each group, or with a scale-free topology according to the algorithm proposed by Barab{\'a}si and Albert \cite{barabasi_s99}. We adopt a systematic approach, going from simple to complex interaction topologies, in order to be able to understand and interpret the results at various levels of interdependence.

For convenience, we denote the two groups as group A and B, and we introduce a probability $p$ according to which each player in group B is allowed to sever one of its in-group links to form a between-group link with one randomly chosen player from group A. The constrain is, however, that no player is allowed to have more than one between-group link. When all players from group B have had the chance to form between-group links, we arrive at the final interaction network consisting of two interdependent groups, whereby the level of interdependence in determined by $p$. The whole procedure is depicted schematically in Fig.~\ref{scheme}. Note that for $p=0$ the two groups are independent, while for $p=1$ they effectively act as one as all the players from group B will be linked with all the players from group A.

After constructing the interdependent groups, we start Monte Carlo simulations of the evolutionary dynamics with uniformly distributed cooperators and defectors, each thus occupying 1/2 of both groups. The accumulation of payoffs $\pi_x$ follows a standard procedure. As noted in the introduction, the prisoner's dilemma game is characterized by the temptation to defect $T=b$, reward for mutual cooperation $R = 1$, and punishment $P$ as well as the sucker's payoff $S$ equalling $0$. Two cooperators facing one another acquire $R$, two defectors get $P$, whereas a cooperator receives $S$ if facing a defector who then gains $T$. The elementary games steps are as follows. First, a player $x$ is randomly selected from either group, and it acquires its payoff $\pi_x$ by playing the game with all its neighbours, including those connected via in-group as well as those connected via between-group links. Next, one randomly chosen neighbour of $x$ within either of the two groups (thus could be connected via an in-group or a between-group link), denoted by $y$, also acquires its payoff $\pi_y$ in the same way. Lastly, if $\pi_y>\pi_x$ player $x$ attempts to adopt the strategy $s_{y}$ from player $y$ with a probability $q=(\pi_y-\pi_x)/(k_{\max}b)$, where $k_{\max}$ is the larger of the two degrees of players $x$ and $y$. This rule is applied instead of the more commonly used Fermi rule \cite{szabo_pre98} to avoid an unequal intensity of selection for players with different degrees.

Presented results were obtained by using a regular topology with $k=6$ for group B, and either a regular, random or scale-free topology with $k=4$ (average degree in latter case) for group A. Players in group B were then allowed to severe one of their in-group links and form a between-group link with probability $p$. The size of individual groups varied from $N=100$ to $10000$, and we confirm the reported results being robust within this interval. For smaller group sizes, however, averages over more independent realizations are needed to obtain statistically consistent simulation results. We have made up to $1000$ independent simulation runs lasting up to $10^5$ full Monte Carlo steps, during each of which all players received the chance once on average to adopt the strategy of one of their neighbours. It is also worth mentioning that exchanging the topologies of groups A and B or varying the degree $k$ does not qualitatively change the presented results and the main conclusions.

\begin{acknowledgments}
We would like to thank Zhi-Xi Wu and Liang Huang for helpful discussions. This research was supported by the National Natural Science Foundation of China (Grants 61203145 and 11047012 to Luo-Luo Jiang) and the Slovenian Research Agency (Grant J1-4055 to Matja{\v z} Perc).
\end{acknowledgments}

\end{document}